\documentclass[aps,prd,amssymb,superscriptaddress,floatfix,nofootinbib,10pt]{revtex4}
\usepackage{graphicx}
\usepackage{subfigure}
\usepackage{hyperref}
\usepackage{amsmath}
\usepackage{hyphenat}
\usepackage{amsfonts}
\usepackage{booktabs}
\usepackage{amssymb}
\usepackage[usenames]{color}
\usepackage{lineno}
\usepackage{bm}
\usepackage{mathptmx}

\usepackage{graphics}
\newcommand{\fm}{\textrm{fm}}

\newcommand{\figcaption}{\def\@captype{figure}\caption}
\newcommand{\tabcaption}{\def\@captype{table}\caption}

\begin{document}

\title{Correlation function for the $T_{bb}$ state: Determination of the binding, scattering lengths, effective ranges and molecular probabilities}

\author{A. Feijoo}
 \thanks{Corresponding author}
	\email{edfeijoo@ific.uv.es}
	\affiliation{Departamento de F\'{\i}sica Te\'orica and IFIC, Centro Mixto Universidad de
	                 Valencia-CSIC Institutos de Investigaci\'on de Paterna, Aptdo.~22085, 46071 Valencia, Spain}

\author{ L.~R.~Dai}
	\affiliation{School of Science, Huzhou University, Huzhou 313000, Zhejiang, China}
	\affiliation{Departamento de F\'{\i}sica Te\'orica and IFIC, Centro Mixto Universidad de
	                 Valencia-CSIC Institutos de Investigaci\'on de Paterna, Aptdo.~22085, 46071 Valencia, Spain}

\author{L.~M.~Abreu}
	\affiliation{Instituto de F\'{\i}sica, Universidade Federal da Bahia, Campus Universit\'{a}rio de Ondina, 40170-115 Bahia, Brazil}
	\affiliation{Instituto de F\'{\i}sica, Universidade de S\~{a}o Paulo, Rua do Mat\~ao, 1371, CEP 05508-090,  S\~{a}o Paulo, SP, Brazil}

\author{E. Oset}
	\email{oset@ific.uv.es}
	\affiliation{Departamento de F\'{\i}sica Te\'orica and IFIC, Centro Mixto Universidad de
       		          Valencia-CSIC Institutos de Investigaci\'on de Paterna, Aptdo.~22085, 46071 Valencia, Spain}

\begin{abstract}
We perform a study of the $B^{*+}B^0,B^{*0}B^+$ correlation functions using an extension of the local hidden gauge approach which provides the interaction from the exchange of light vector mesons and gives rise to a bound state of these components in $I=0$ with a binding energy of about $21$~MeV. After that, we face the inverse problem of determining the low energy observables, scattering length and effective range for each channel, the possible existence of a bound state, and, if found, the couplings of such a state to each $B^{*+}B^0,B^{*0}B^+$ component as well as the molecular probabilities of each of the channels. We use the bootstrap method to determine these magnitudes and find that, with errors in the correlation function typical of present experiments, we can determine all these magnitudes with acceptable precision. In addition, the size of the source function of the experiment from where the correlation functions are measured can be also determined with a high precision.
\end{abstract}

\date{\today}

\maketitle

\section{Introduction}
The study of correlation functions in pairs of particles observed in high energy $p-p$, $p-A$ and $A-A$ collisions is turning into a very useful tool to determine the basic properties of the pair interaction \cite{Lisa:2005dd,ALICE:2020mfd,ALICE:2022wwr,Fabbietti:2020bfg}. Experimental work in the strangeness sector is abundant \cite{ALICE:2020mfd,ALICE:2018ysd,ALICE:2019gcn,ALICE:2021njx,ALICE:2019buq,ALICE:2019eol,ALICE:2019hdt,ALICE:2021cpv,ALICE:2021cyj}, but the ALICE collaboration is starting to explore the charm sector measuring correlation functions in high-multiplicity $pp$ reactions at $13$~TeV \cite{ALICE:2022enj}. In the future one will also have access to the bottom sector.

Theoretically the correlation functions are obtained from the overlap of the source function with the square of the wave function of the produced pair \cite{Koonin:1977fh,Lednicky:1981su,Pratt:1986cc,Pratt:1990zq,Bauer:1992ffu,Morita:2014kza,Ohnishi:2016elb,Morita:2016auo,Hatsuda:2017uxk,Mihaylov:2018rva,Haidenbauer:2018jvl,Morita:2019rph,Kamiya:2019uiw,Kamiya:2021hdb,Kamiya:2022thy,Vidana:2023olz,Liu:2023uly,Albaladejo:2023pzq,Torres-Rincon:2023qll}. While most of the theoretical studies simply compare the results of a model with the experimental correlation functions, it has only been recently that the inverse problem of obtaining the observables related to the interaction of the pair has been faced. In this sense, in Ref.~\cite{Ikeno:2023ojl} it was shown that from the knowledge of the $D^0K^+$, $D^+K^0$ and $D_s^+\eta$ correlation functions one could determine the existence of a $D_{s0}$ bound state, which corresponds to the $D^*_{s0}(2317)$, the scattering lengths and effective ranges of the channels, and the molecular probabilities of the state in each of the channels, together with the extent of the source. A thorough analysis using the bootstrap method provided the errors in the determination of these observables given a certain precision in the experimental data. The same was done in the investigation of the correlation functions of $D^{*+}D^0$ and $D^{*0}D^+$ in Ref.~\cite{Albaladejo:2023wmv}, from where one could conclude that knowing the correlation functions one could obtain the existence of the $T_{cc}$ state \cite{LHCb:2021vvq,LHCb:2021auc}, its binding, width, scattering lengths and effective ranges, size of the source, as well as the molecular probability of the state, all this with relatively high precision, assuming errors in the correlation functions as the present ones in other reactions.

While in principle the information obtained from the correlation functions should be equivalent to that obtained from mass distributions in analogous experiments, the fact that the source function changes from the $p-p$, $p-A$ or $A-A$ experiments, adds extra information which is most valuable to extract information on the interaction of the studied pairs.

The $T_{bb}$ state, analogous to the $T_{cc}$ in the bottom sector, would be built up from the $B^{*+}B^0$ and $B^{*0}B^+$ channels and was predicted in \cite{Dai:2022ulk} as a molecular state of these components with Isospin $I=0$, a binding of $21$~MeV and a width of $14$~eV, coming from the radiative decay of the $B^*$. Actually, this state is also investigated from different perspectives \cite{Tornqvist:1993ng,Ohkoda:2012hv,Li:2012ss,Liu:2020nil,Manohar:1992nd,Zhao:2021cvg,Ke:2021rxd,Yang:2009zzp,Barnes:1999hs,Michael:1999nq,Detmold:2007wk,Brown:2012tm,Bicudo:2012qt,Bicudo:2015kna,Carlson:1987hh,Bicudo:2016ooe,SanchezSanchez:2017xtl,Wang:2018atz,Yu:2019sxx,Ding:2020dio,Ding:2021igr,Meng:2020knc,Abreu:2022sra,Ader:1981db,Hernandez:2019eox,Francis:2016hui,Junnarkar:2018twb,Leskovec:2019ioa,Mohanta:2020eed,Hudspith:2023loy,Aoki:2023nzp} which conclude the existence of this state (see Ref.~\cite{Dai:2022ulk} for a detailed information of these works).

In the present work we start from the picture of \cite{Dai:2022ulk} and construct the two correlation functions for $B^{*+}B^0$ and $B^{*0}B^+$ and then proceed with the inverse problem of determining the different observables that can be extracted from the correlation function. The method is very general and does not assume the state to be molecular, some freedom is left for the contribution of non-molecular components. Certainly, while starting from the picture of \cite{Dai:2022ulk} the answer cannot be other than the input of that theoretical framework, the important result is the errors induced in the observables related to that state, assuming errors in the correlation function of the size in the present measurements. We find that all these observables can be obtained with certain accuracy, including the size of the source function. This means that a measurement of these correlation functions can provide the binding of the state, the scattering lengths and effective ranges of the $B^{*+}B^0$ and $B^{*0}B^+$ channels, the molecular probabilities and the isospin nature of the state, in addition to the size of the source function.

\section{Formalism}
We first recall the details on how the $T_{bb}$ state is obtained in \cite{Dai:2022ulk} and then construct the correlation functions. In a last step, we discuss the inverse problem of getting observables from the correlation functions.

\subsection{Model for $B^{*+}B^0,B^{*0}B^+$ interaction}

In \cite{Dai:2022ulk} we use an extension of the local hidden gauge approach \cite{Bando:1987br,Harada:2003jx,Meissner:1987ge,Nagahiro:2008cv} where the vector mesons are exchanged in the interaction of $B^*B$. We obtained an interaction given by
\begin{widetext}
\begin{eqnarray}\label{eq:prd}
\hspace{3.cm}&V = C_{ij} \, g^2\, \frac{1}{2} \left[3s-(M^2+m^2+M'^2+m'^2) -\frac{1}{s}(M^2-m^2)(M'^2-m'^2) \right] \,,
\end{eqnarray}
\end{widetext}
where $M, M'$ are the initial, final vector masses and $m, m'$ the initial, final pseudoscalar masses present in the corresponding state in the two channel formalism with $B^{*+}B^0 \, (1)$ and $B^{*0}B^+ \, (2)$. Their threshold masses are very similar, being $10604.37$~MeV for $B^{*+}B^0$ and  $10604.96$~MeV for $B^{*0}B^+ $.  The coefficients $C_{ij}$ are given by the matrix
\begin{eqnarray}\label{eq:cij}
C_{ij} = \left(
           \begin{array}{cc}
            0  & \frac{1}{m_\rho^2}    \\[0.1cm]
           \frac{1}{m_\rho^2}   &   0 \\
           \end{array}
         \right)\,,
\end{eqnarray}
and $g=M_V/2f$ ($M_V=800$~MeV,$f=93$~MeV). The diagonal terms in the $C_{ij}$-matrix are zero because we neglect $1/M^2_{\Upsilon}$ versus $1/M^2_{\rho}$. One cannot expect any bound state for $B^{*+}B^0$ or $B^{*0}B^+ $ considered as single channels, but if one diagonalizes $C_{ij}$ one finds two eigenstates that correspond to the isospin states
\begin{eqnarray}\label{eq:iso}
  |B^* B, I=0\rangle = -\frac{1}{\sqrt{2}} (B^{*+} B^0 - B^{*0} B^+)\,, &  |B^* B, I=1\rangle =- \frac{1}{\sqrt{2}} (B^{*+} B^0 + B^{*0} B^+)\,.
\end{eqnarray}
One can see that for these combinations one finds $C(I=0)=-\frac{1}{m_\rho^2} $; $C(I=1)=\frac{1}{m_\rho^2} $ indicating attraction in $I=0$ and repulsion in $I=1$. In  \cite{Dai:2022ulk} one finds that the $I=0$ component develops a bound state with this potential. This is seen by looking at the poles of the $T$ given by
\begin{eqnarray}\label{eq:BSeq}
T=[1-VG]^{-1} \, V \, ,
\end{eqnarray}
with $G$ the diagonal loop function for $B^*B$ propagation, regularized with a cutoff, $q_{\rm max}$, and given by 

\begin{align}\label{eq:Gcut}
\hspace{-0.25cm}&G = \int_{|{\vec q}|<q_{\rm max}} \frac{d^3q}{(2\pi)^3} \, \frac{\omega_1 (q)+ \omega_2(q)}{2 \,\omega_1(q)  \omega_2(q)} \,\frac{1}{s-(\omega_1(q) + \omega_2(q))^2+i\epsilon}
\end{align}
where $\omega_i(q)=\sqrt{m_i^2+{\vec q}^2}$. The value of $q_{\rm max}$ used in \cite{Dai:2022ulk} is $420$~MeV which we shall also use here. Eq.~(\ref{eq:BSeq}) with the cutoff regularization can be justified using dispersion relations \cite{Oller:2000fj}, but can be equally obtained using a separable potential $V(\vec q,\vec q')=V \theta(q_{\rm max}-|\vec q|)\theta(q_{\rm max}-|\vec q'|)$ which leads to a $T$ matrix with the same structure $T(\vec q,\vec q')=T \theta(q_{\rm max}-|\vec q|)\theta(q_{\rm max}-|\vec q'|)$, which we take into account when evaluating the correlation functions in \cite{Vidana:2023olz}.

\subsection{Correlation functions}

Following the approach of \cite{Vidana:2023olz} we write the correlation functions
\begin{align}
\hspace{-0.25cm}&C_{B^0B^{*+}}(p_{B^0})=1+4\pi\,\theta(q_\text{max}-p_{B^0}) \int_0^{+\infty} drr^2S_{12}(r)\Big\{\big|j_0(p_{B^0}r)+T_{11}(E)\widetilde{G}^{(1)}(r;E)\big|^2 
+\big|T_{21}(E)\widetilde{G}^{(2)}(r;E)\big|^2-j^2_0(p_{B^0}r)\Big\} \, ,
\label{eq:cf4}
\end{align} 
with $E=\sqrt{s}$ and  being 
\begin{equation}
p_{B^0}=\frac{\lambda^{\frac{1}{2}}(s,m_{B^0}^2,m_{B^{*+}}^2)}{2\sqrt{s}} \, ; \nonumber \\
\end{equation}
\begin{align}
\hspace{-0.25cm}&C_{B^+B^{*0}}(p_{B^+})=1+4\pi\,\theta(q_\text{max}-p_{B^+}) 
\int_0^{+\infty} drr^2S_{12}(r)\Big\{\big|j_0(p_{B^+}r)+T_{22}(E)\widetilde{G}^{(2)}(r;E)\big|^2 
+\big|T_{12}(E)\widetilde{G}^{(1)}(r;E)\big|^2-j^2_0(p_{B^+}r)\Big\} \, ,
\label{eq:cf5}
\end{align} 
with 
\begin{equation}
p_{B^+}=\frac{\lambda^{\frac{1}{2}}(s,m_{B^+}^2,m_{B^{*0}}^2)}{2\sqrt{s}} \, , \nonumber \\
\end{equation}
where $T_{ij}$ are the scattering matrices obtained from Eq.~(\ref{eq:BSeq}) and the $\widetilde{G}^{(i)}$ function is given by
\begin{align}
\widetilde{G}^{(i)}(r;E)&=\int_{|\vec q\,|< q_\text{max}} \frac{d^3\vec q}{(2\pi)^3}\frac{\omega^{(i)}_1(q)+\omega^{(i)}_2(q)}{2\omega^{(i)}_1(q)\,\omega^{i}_2(q)}
\frac{j_0(qr)}{s-\left[\omega_1^{(i)}(q)+\omega_2^{(i)}({q})\right]^2+i\epsilon}
\label{eq:Gtilde}
\end{align} 
with the source function 
\begin{equation}
S_{12}(r) = \frac{1}{(\sqrt{4\pi})^3 R^3}\mbox{exp}\left(-\frac{r^2}{4R^2}\right) \ .
\label{sec:source}
\end{equation}
%
\begin{figure}[t]
\begin{center}
\includegraphics[width=0.55\textwidth,keepaspectratio]{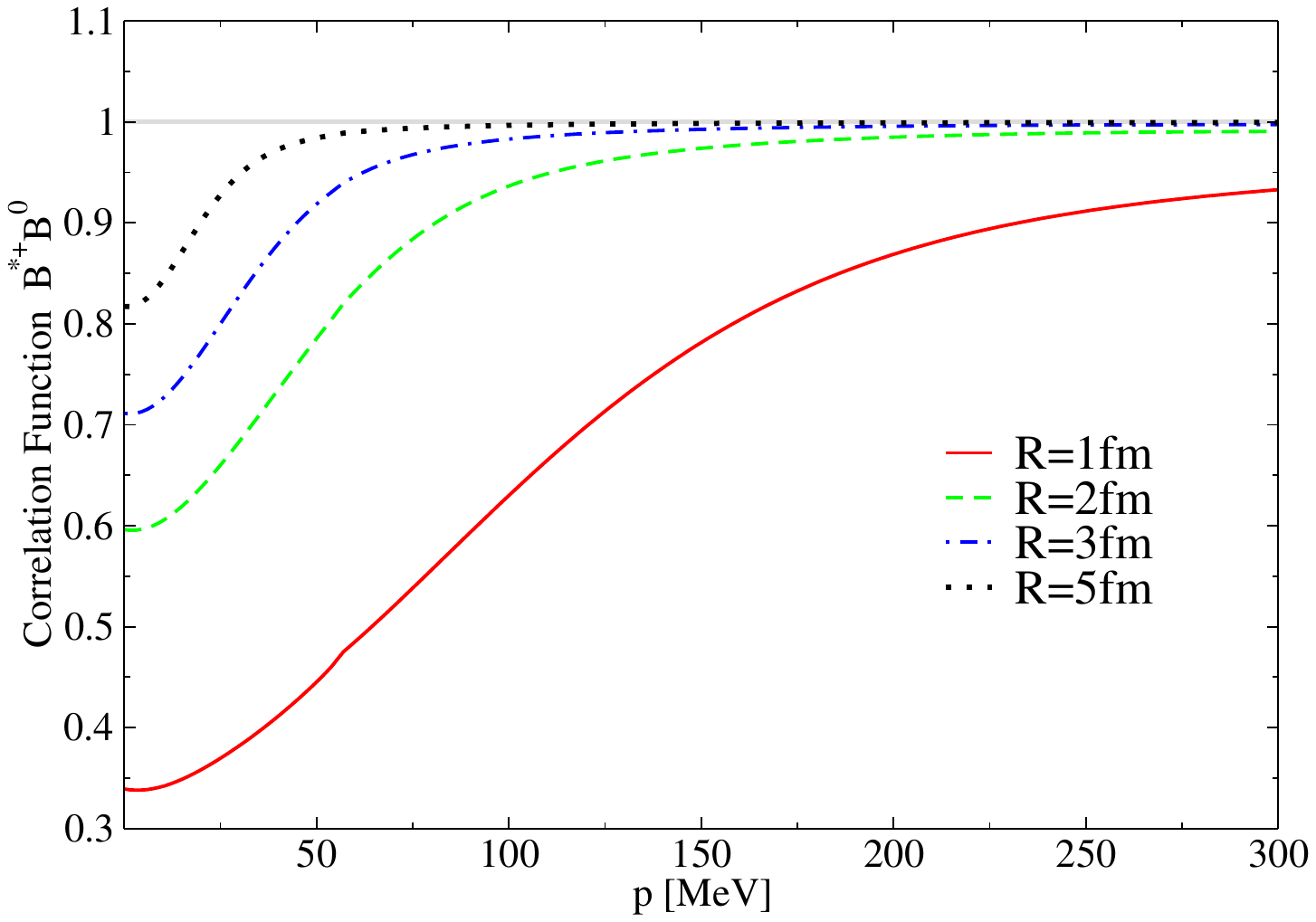}
\caption{(Color online). Correlation function of the $B^0B^{*+}$ pair for different values of the source size ($R$) with fixed $q_{\rm max}= 420$~MeV.}
\label{fig:fig1}
\end{center}
\end{figure}
\begin{figure}[t]
\begin{center}
\includegraphics[width=0.55\textwidth,keepaspectratio]{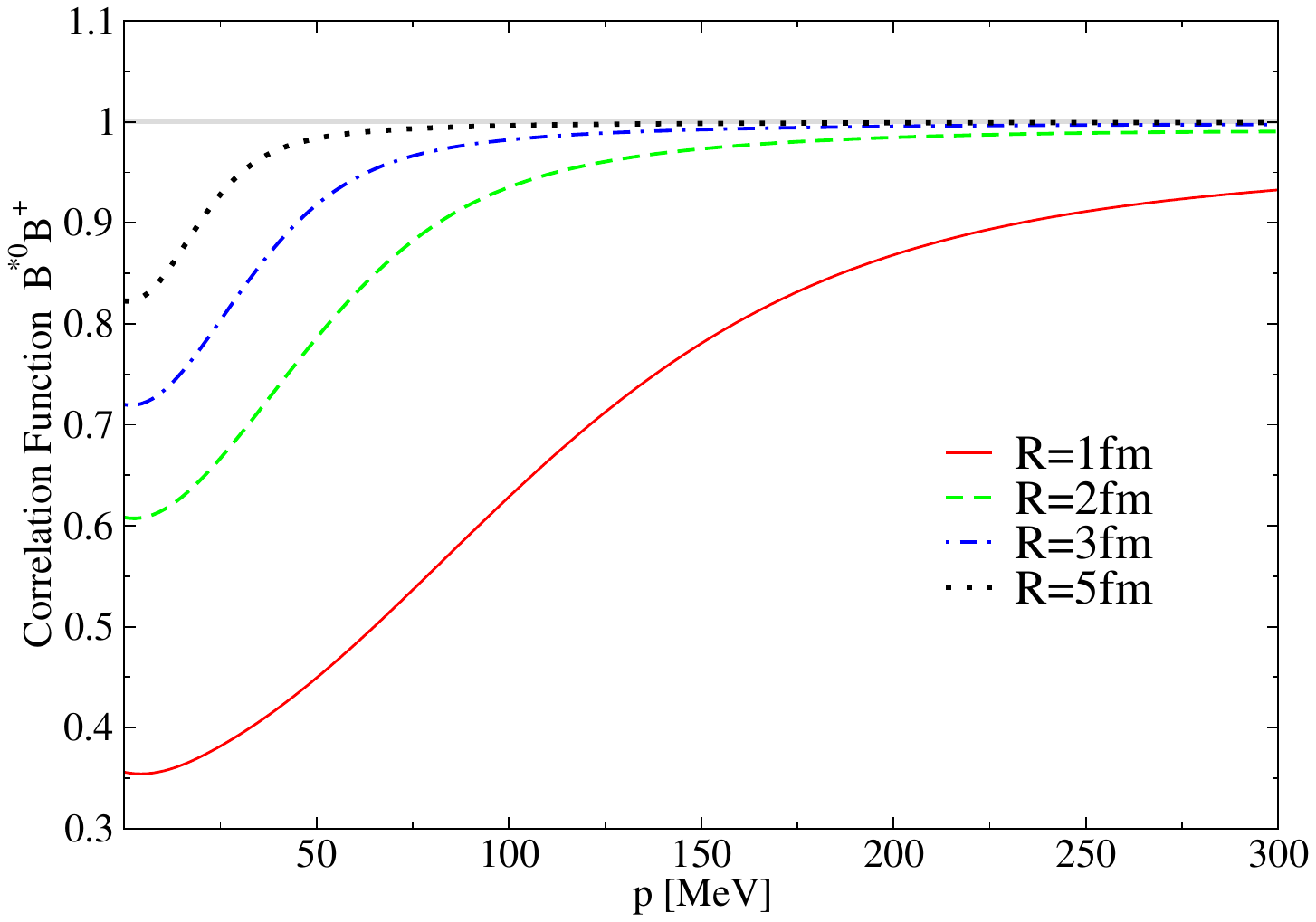}
\caption{(Color online). Correlation function of the $B^+B^{*0}$ pair for different values of the source size ($R$) with fixed $q_{\rm max}= 420$~MeV.}
\label{fig:fig2}
\end{center}
\end{figure}

\subsection{Inverse problem}

To do the inverse problem we start from the generated correlation functions and produce synthetic data, choosing $31$ points from each correlation function with a homogeneous error corresponding to the $10\%$ of the minimum value taken by the  correlation function. Then we assume a potential
\begin{eqnarray}\label{eq:v1}
V = \left(
           \begin{array}{cc}
            V_{11} & V_{12}   \\[0.1cm]
           V_{12} &  V_{22}\\
           \end{array}
         \right)\,.
\end{eqnarray}

Next, we assume that the potential has isospin symmetry, $\langle I=0 \vert V  \vert  I=1 \rangle = 0$, not assuming any particular isospin. This implies that $V_{11}=V_{22}$, and hence the matrix $V$ becomes
\begin{eqnarray}\label{eq:v2}
V = \left(
           \begin{array}{cc}
            V_{11} & V_{12}   \\[0.1cm]
           V_{12} &  V_{11}\\
           \end{array}
         \right)\,.
\end{eqnarray}

We also allow for the possibility that there is some contribution from nonmolecular states, which is done introducing some energy dependent terms as done in \cite{Ikeno:2023ojl,Dai:2023cyo}, following the results obtained in \cite{Hyodo:2013nka,Aceti:2014ala} to absorb the effects of eliminated channels into those which are kept. We take thus, 
\begin{eqnarray}\label{eq:vv}
V_{11} = V'_{11} + \frac{\alpha}{m^2_V} (s-s_0) \,, & V_{12} = V'_{12} + \frac{\beta}{m^2_V} (s-s_0)\,,
\end{eqnarray}
with $s_0$ the energy squared of the lowest threshold, $B^{*+}B^0$. In Eq.~(\ref{eq:vv}), $\alpha$ and $\beta$ are free parameters and $m^2_V$ is introduced as a scale to have the $\alpha$ and $\beta$ parameters dimensionless.

We then carry the best fit to the synthetic data and determine the parameters $q_{\rm max},V'_{11},V'_{12},\alpha,\beta,R$. There are certainly correlations between these parameters and different sets produce the same results. One can see, anticipating that we have an $I=0$ state, that the relevant parameters are $V'_{11}-V'_{12}$ and $\alpha-\beta$. There are also some correlations between $q_{\rm max}$ and $V'_{11},V'_{12}$. This simply means that the values obtained for the parameters are not very meaningful. The relevant information is the value of the observables, and to get them and, very important, their uncertainties, we use the resampling (bootstrap) method \cite{PresTeukVettFlan92,Efron:1986hys,Albaladejo:2016hae} to determine these magnitudes. For this, we generate with a Gaussian distribution the centroids of the synthetic data chosen and make a fit to these data, obtaining values of the parameters from where the magnitudes are determined. The procedure is iterated a number of times, about $50$, and after that the average value of each observable is evaluated as well as its dispersion. One restriction is put in the parameters: The formulas for $T_{11},T_{22}$ and $T_{12}^2$ depend on $V_{12}^2$ and hence the sign of $V_{12}$ cannot be determined. To avoid a misinterpretation of the results, we take a minimal information from the local hidden gauge approach which is that $V'_{11}-V'_{12}$ should be negative, as it comes unquestionably from Eq.~(\ref{eq:cij}). Another technical question, which saves computing time and gives precision, is that in Eq.~(\ref{eq:Gtilde}) we separate
\begin{equation}
j_0(qr)=j_0(qr)-j_0(q_{\rm on}r)+j_0(q_{\rm on}r)
\end{equation}
with $q_{\rm on}$ the on shell value of the momentum in the loop 
\begin{equation}
q_{\rm on}=\frac{\lambda^{\frac{1}{2}}(s,m_{1}^2,m_{2}^2)}{2\sqrt{s}} \, .\nonumber \\
\end{equation}
The part of $j_0(qr)-j_0(q_{\rm on}r)$ cancels the pole of $\widetilde{G}$ and the term with $j_0(q_{\rm on}r)$ can be calculated analytically from a formula in \cite{Oller:1998hw} (see details in \cite{Albaladejo:2023wmv}). The expressions to obtain the scattering lengths, effective ranges, couplings and probabilities are identical to those used in Ref.~\cite{Dai:2023cyo} and we refrain from reproducing them here.

\section{Results}
\label{sec:results}

In Figs. \ref{fig:fig1} and \ref{fig:fig2} we plot the results of the correlation functions of $B^0B^{*+}$ and $B^+B^{*0}$ for different values of the range parameter $R$ of the source. We can see that the size of the correlation functions changes appreciably with $R$. On the other hand, given the proximity of the two thresholds, the correlation functions are remarkably similar. 

In order to discern which could be the goodness and reliability of the information one can extract from the correlation functions depending on the primary colliding elements ($p-p$, $A-A$), we perform two different studies. First we proceed with the analysis of synthetic data generated employing a source size $R^{\rm input}= 1 {\rm fm }$, which would correspond to mimic a $p-p$ collision, then we iterate the process taking into account a second set of synthetic data generated from a source with $R^{\rm input}= 5 {\rm fm }$ ($A-A$ collision). We obtain these two sets of parameters:
\begin{eqnarray}\label{param_fit}
R^{\rm input}= 1 {\rm fm }: \,\,\, q_{max} = 445 \pm 29 \, {\rm MeV} ,  &  V'_{11} = 70 \pm 360  ,  &  V'_{12} = 3463 \pm 1272 ,      \nonumber \\
\alpha = -170 \pm 336 & \beta =  290 \pm 346  & R = 0.98 \pm 0.02  \, {\rm fm} \, ; \nonumber \\
R^{\rm input}= 5 {\rm fm }: \,\,\, q_{max} = 402  \pm 93 \, {\rm MeV} ,  &  V'_{11} = -500 \pm 376  ,  &  V'_{12} = 3567 \pm 1779,      \nonumber \\
\alpha = -292  \pm 347 & \beta = 496  \pm 348  & R = 4.99 \pm 0.61  \, {\rm fm} \, .
 \end{eqnarray}
 
 \begin{table*}[ht!]
\centering
\renewcommand\arraystretch{1.5}
\caption{\label{tab:tab1} The obtained scattering lengths and effective ranges for both bootstrap analysis.}
\begin{tabular}{ccccccc}
\hline\hline
 & $R^{\rm input} [{\rm fm }]$ & $a_1~[\fm]$ & $r_{0,1}~[\fm]$ &  $a_2~[\fm]$ & $r_{0,2}~[\fm]$ & \\
\hline
 &$ 1 $ & $0.85 \pm 0.18$ & $-0.11 \pm 0.51$ & $ (0.81\pm 0.13) -i \, (0.03 \pm 0.03)$ & $(0.43 \pm 0.11)-i\, (0.38 \pm 0.29)$  &\\
 &$ 5 $ & $0.85 \pm 0.19$ & $-0.92 \pm 1.78$ & $ (0.77\pm 0.13) -i \, (0.05 \pm 0.06)$ & $(0.26 \pm 0.40)-i\, (0.87 \pm 1.13)$  &\\

\hline\hline
\end{tabular}
\end{table*}

\begin{table*}
\centering
\renewcommand\arraystretch{1.5}
\caption{\label{tab:tab2} The obtained coupling constants and probabilities.}
\begin{tabular}{cccccccccc}
\hline\hline
 & $R^{\rm input} [{\rm fm }]$ &  $g_1$ [MeV] & $g_2$ [MeV] &  $P_1$ & $P_2$ &  $Z$\\
\hline
& 1 & $33039 \pm 14744$  & $-32031 \pm 17367 $  &  $0.44 \pm 0.06$ & $0.43 \pm 0.05$ & $0.13\pm 0.11$\\
& 5 & $ 30970 \pm 19666$  & $-31181 \pm  19718$  &  $ 0.41\pm 0.11$ & $0.39 \pm 0.11$ & $0.19\pm 0.22$\\

\hline\hline
\end{tabular}
\end{table*}
\begin{figure}[t]
\begin{center}
\includegraphics[width=0.75\textwidth,keepaspectratio]{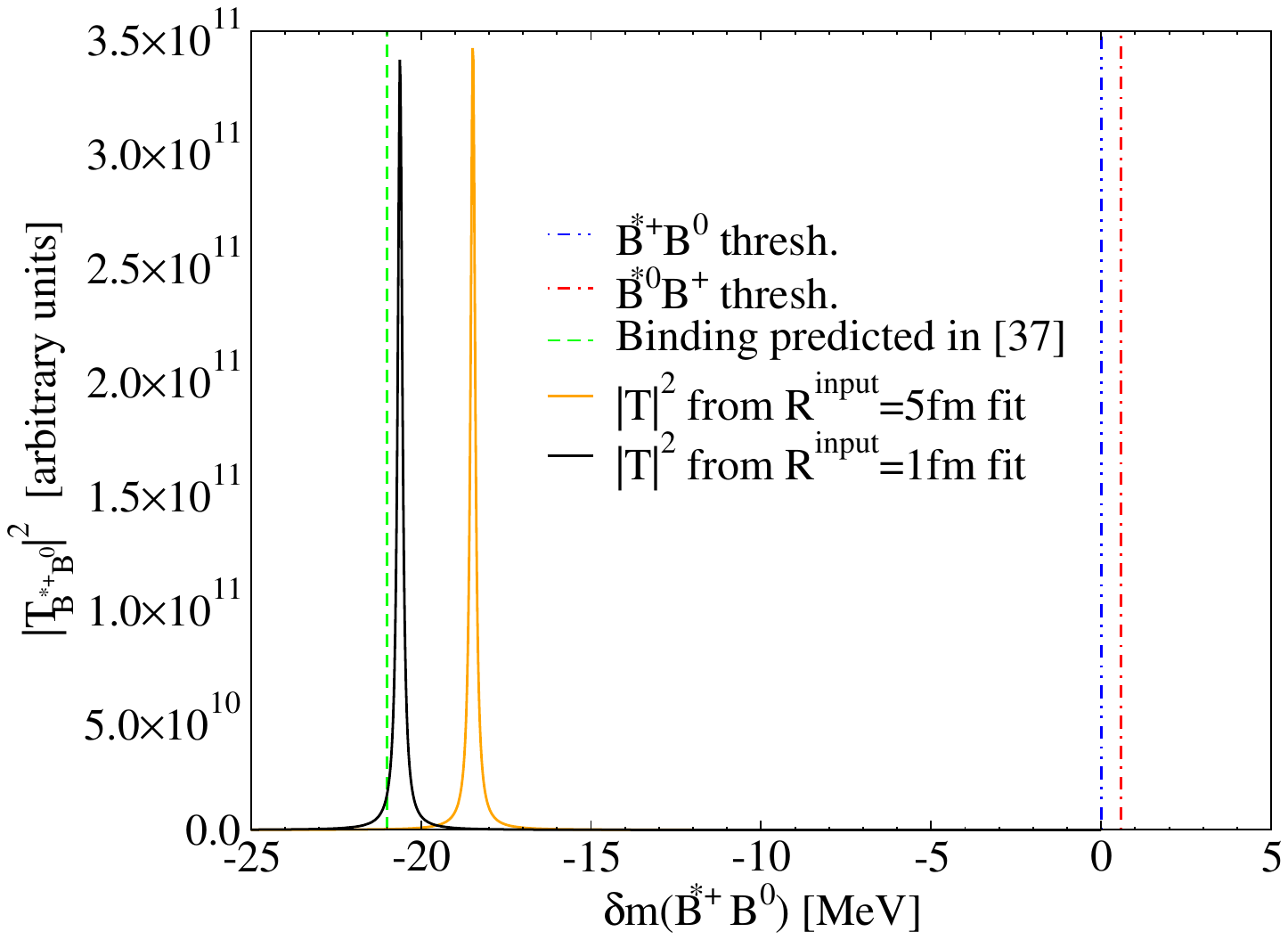}
\caption{(Color online). The obtained modulus squared of the amplitudes, $\vert T_{B^0B^{*+},B^0B^{*+}}\vert^2$, employing the best-fit parameters displayed in Eq.~(\ref{param_fit}).}
\label{fig:fig3}
\end{center}
\end{figure}
With these parameters we obtain both $T_{B^0B^{*+},B^0B^{*+}}$ amplitudes and their modulus squared are plotted in Fig. \ref{fig:fig3}. We get two very narrow peaks corresponding to bound states (the tiny width of the $B^*$ is neglected in the calculations). The binding energies obtained from the best fits are $20.62$~MeV ($R^{\rm input}= 1 {\rm fm }$) and $18.48$~MeV ($R^{\rm input}= 5 {\rm fm }$), in remarkably good agreement with the resulting $21$~MeV found in \cite{Dai:2022ulk}. This already shows the value of the analysis done, which allows from the structureless correlation functions of Figs. \ref{fig:fig1} and \ref{fig:fig2} to deduce that there is a bound state of the $B^0B^{*+}$ (and $B^+B^{*0}$) system.

The values of the parameters in Eq.~(\ref{param_fit}) give us a feeling of their strength, but as discussed  before, these particular values are not meaningful given the correlations between the parameters. In order to obtain the values of the observables and their uncertainties we use the bootstrap method and run $50$ best fits with the resampled data. In each of the fits we determine their average and their dispersion. The results are summarized in Tables \ref{tab:tab1} and \ref{tab:tab2}. The scattering lengths are determined with a $20\%$ precision while, for the components of the effective ranges, we get larger uncertainties. This larger error is understandable since the contribution of the effective range in the T matrix in the range of the correlation functions is smaller than that of the scattering length. In addition, since the evaluation of the effective range requires a derivative with respect to $s$ (see Eq~(26) in \cite{Dai:2023cyo}), it involves the terms with $\alpha$ and $\beta$, and their fluctuations in the resampling method add to the final uncertainty of this magnitude. Since the information of the effective range is more important when we move away from threshold, it is not surprising that we also get a larger relative error in the determination of the binding and the couplings. Despite the previous fact, the average values provided by the bootstrap method for both cases are in remarkable agreement with the results shown in \cite{Dai:2022ulk} as can be seen in Table \ref{tab:tab2} and in the values obtained for the bindings and their uncertainties 
\begin{equation}
R^{\rm input}= 1 {\rm fm }: B=-22 \pm 15 \,\, {\rm MeV} \, ; \,\,\, R^{\rm input}= 5 {\rm fm }: B=-22 \pm 21 \,\, {\rm MeV} \, .
\end{equation}
It is even more interesting to note that $g_1$ and $g_2$ are very similar with opposite sign in both analysis, which indicates from Eq.~(\ref{eq:iso}) that they correspond to a state of $I=0$.

We come back here to the issue of the restriction to take $V'_{11}-V'_{12}$ negative. For this we recall Eqs.~(17), (18) of Ref.~\cite{Dai:2023cyo}, which gives for two channels (ignore $\alpha$ and $\beta$ for the discussion)
\begin{eqnarray}\label{eq:Tdis}
T = \frac{1}{DET}\left(
           \begin{array}{cc}
           V_{11}+(V_{12}^2-V_{11}^2)G_2  & V_{12}   \\[0.1cm]
           V_{12} &  V_{11}+(V_{12}^2-V_{11}^2)G_1   \\
           \end{array}
         \right)\,.
\end{eqnarray}
with $DET$ the determinant of $1-VG$, given by $DET=1-V_{11}(G_1+G2)-(V_{12}^2-V_{11}^2)G_1G_2$.
We can see that $T_{11},T_{22}$ contain quadratic terms in $V_{12}$, nevertheless, $T_{12}$ contains quadratic terms in $V_{12}$ but is also proportional to $V_{12}$. However, in the correlation functions of Eqs.~(\ref{eq:cf4}),(\ref{eq:cf5}) $T_{12}$ appears quadratic, which means that from the information of the correlation functions we can only obtain $V_{12}^2$, and the sign of $V_{12}$ is not determined. If we go back to Eq.~(\ref{param_fit}) we can then see that a solution with $V'_{11}=70$ and $V'_{12}=-3463$ would be equally acceptable. Since the couplings $g_i$ are defined as 
\begin{equation}
g_1^2=\lim_{s \to s_0}(s-s_0)T_{11}, \,\,\,\, g_2=\lim_{s \to s_0}(s-s_0) \frac{g_1T_{12}}{T_{11}}
\label{defcoup}    
\end{equation}
with $s_0$ the square of the bound-state mass, a change of sign in $V'_{12}$, and hence in $T_{12}$, implies a change of sign of $g_2$. Then in Table~\ref{tab:tab2} we would get the couplings $g_1,g_2$ about the same and with positive sign. According to Eq.~(\ref{eq:iso}) this would mean that we would have a state with $I=1$ coming from the potential $V'_{11}+V'_{12}$, which is now attractive. Hence, technically this solution is possible in the present analysis. But here we invoke the combination of two elements: On one side, an experimental fact, which is that the $T_{cc}$ was found in \cite{LHCb:2021auc} to be a state of $I=0$, and no signal was found in the $I=1$ channel. Next we invoke heavy quark flavor symmetry \cite{Manohar:1992nd}, which would make us conclude that there must be a state of $I=0$ also for the $T_{bb}$. Based on this argumentation we impose that $V'_{11}-V'_{12}$ should be negative to prevent the interpretation of the results as having an $I=1$ state. Even then, one could obtain $g_1$, $g_2$ with opposite sign but not too close to each other in absolute value. The results of Table~\ref{tab:tab2}, within errors, indicate that we have an $I=0$ state rather clean.

The probabilities of having $B^0B^{*+}$ and $B^+B^{*0}$, $P_1$, $P_2$, are very close to $0.5$ each, in both cases, and their sum is compatible with $1$ within errors ($P_1+P_2=0.87 \pm 0.11$) and ($P_1+P_2=0.80 \pm 0.2$) for $R^{\rm input}= 1 {\rm fm }$ and $R^{\rm input}= 5 {\rm fm }$ respectively, indicating that the nature of the state is mainly molecular. Actually, the nonmolecular strength $Z=1-(P_1+P_2)$ is $0.13$ for $R^{\rm input}= 1 {\rm fm }$ with an error  $\pm 0.11$  and $0.13 \pm 0.22$ for $R^{\rm input}= 5 {\rm fm }$ that make them basically compatible with zero. 

To conclude, one of the most important results of the present study is that one can obtain values for both sources as
\begin{eqnarray}
& R^{\rm input}= 1 {\rm fm }: \,\, R=0.974 \pm 0.024 \,\, {\rm fm} \,  \nonumber \\
& R^{\rm input}= 5 {\rm fm }: \,\, R=5.052 \pm 0.614 \,\, {\rm fm} \, .   
\end{eqnarray}
with a notable precision comparing them to the starting input of $R=1\, {\rm fm} $ and $R=5\, {\rm fm} $ used to generate the synthetic data.

In Ref.~\cite{Dai:2022ulk}, in addition to the $BB^{*}(I=0)$ system, other related states were also found: the $BB^{*}_s-B_sB^{*}(I=1/2)$, $B^{*}B^{*}(I=0)$ and $B^{*}_sB^{*}(I=1/2)$, all of them with $J^P=1^+$. The formalism developed here could be extended automatically to the $BB^{*}_s, B_sB^{*}(I=1/2)$ channels. The other two states decay to $BB^{*}$ or $BB^{*}_s, B_sB^{*}$ and dealing with these cases could also be done, adding the decay channels as new coupled channels or using complex potentials, which would make the formalism a bit more complicated.

\section{Conclusion}
\label{sec:conclusions}

Using input from the local hidden gauge approach, exchanging vector mesons between the vector and pseudoscalar in the $B^0B^{*+}$ and $B^+B^{*0}$ systems as a source of their interaction, we have evaluated the scattering matrices and the correlation functions of the two systems. After that, we have addressed the inverse problem of determining the low energy observables related to this interaction from the knowledge of the correlation functions. For this, we have parametrized the potential in a very general form which does not preassume the existence of a bound state nor its nature in the case of existence. The formalism explicitly has the freedom to incorporate the effect of some state of nonmolecular nature.

Very important in the study is to determine the errors of the observables assuming errors in the correlation function data of the size obtained in present experiments. We observe that the correlation functions are rather smooth and do not indicate a priori that they are linked to the presence of a bound state, but we could find a bound state about $22$~MeV below threshold in agreement with the theory used to construct the correlation functions. It is by no means trivial that the information of the correlation functions, which spans for an energy above threshold of about $17$~MeV can provide precise information about the amplitudes $20$~MeV below threshold, but this is what is found. Actually, as counterpart, the extraction of such information incorporates uncertainties ranging from $10\%$ to  $40\%$ for those magnitudes related to the bound state. Despite this, it can be clearly appreciated that the couplings of the bound state obtained for $B^0B^{*+}$ and $B^+B^{*0}$ have very similar size and opposite sign thereby characterizing this state as a pure $I=0$ state. With the value of these couplings we could also calculate the molecular probabilities of these components in the wave function of the state, which were around $0.44$ each (almost compatible to $0.5$ within errors), indicating a clear molecular nature for the state obtained. This is not surprising, since this was assumed in the model used to construct the correlation functions, but the novelty here is that we use, in the analysis, tools that allow for the contribution of nonmolecular states, such that if the $T_{bb}$ state was of a different nature than assumed, and the correlation functions different than those constructed here under the molecular assumption, the method can give us the probability of any nonmolecular component.

The situation improves for both scattering lengths since these observables are evaluated at the corresponding thresholds where the terms with $\alpha$ and $\beta$ are either $0$ or take a small value by construction, while the effective ranges acquire higher error values as we have discuss above. 

Finally, and also not trivial, we found that the inverse method used in the analysis can also provide the size of the source function with great precision. We performed the analysis of the inverse method for two source sizes, $R=1 {\rm fm }$ and $R=5 {\rm fm }$. In the latter case, the correlation function has a smaller diversion from unity and we would expect that the inverse method provides a smaller precision for the observables. Even then we find results remarkably similar in both cases, compatible within in errors, and we observed that we can also get the size of the source in both cases, only, in the case of $R=1 {\rm fm }$ the error is of the order of $2.5\%$, while for $R=5 {\rm fm }$ the uncertainty is about the $12\%$.

All these findings can only encourage experimental groups to look for these correlation functions with the certainty that one can extract valuable information for a large number of observables with acceptable precision.

\section{acknowledgement}
This work was supported by the Spanish Ministerio de Ciencia e Innovaci\'on (MICINN) and European FEDER funds under Contracts No.\,PID2020-112777GB-I00, and by Generalitat Valenciana under contract PROMETEO/2020/023. This project has received funding from the European Union Horizon 2020 research and innovation programme under the program H2020-INFRAIA-2018-1, grant agreement No.\,824093 of the STRONG-2020 project. A.\,F. is supported through Generalitat Valencia (GVA) Grant APOSTD-2021-112. This work is partly  supported by the National Natural Science Foundation of China under Grants Nos. 12175066, 11975009. The work of L.M.A. is partly supported by the Brazilian agencies CNPq (Grant Numbers 309950/2020-1, 400215/2022- 5, 200567/2022-5),  and CNPq/FAPERJ under the Project INCT-F\'isica Nuclear e Aplica\c{c}\~oes (Contract No. 464898/2014-5).


\bibliography{refs.bib}

\begin{thebibliography}{80}
\expandafter\ifx\csname natexlab\endcsname\relax\def\natexlab#1{#1}\fi
\expandafter\ifx\csname bibnamefont\endcsname\relax
  \def\bibnamefont#1{#1}\fi
\expandafter\ifx\csname bibfnamefont\endcsname\relax
  \def\bibfnamefont#1{#1}\fi
\expandafter\ifx\csname citenamefont\endcsname\relax
  \def\citenamefont#1{#1}\fi
\expandafter\ifx\csname url\endcsname\relax
  \def\url#1{\texttt{#1}}\fi
\expandafter\ifx\csname urlprefix\endcsname\relax\def\urlprefix{URL }\fi
\providecommand{\bibinfo}[2]{#2}
\providecommand{\eprint}[2][]{\url{#2}}

\bibitem[{\citenamefont{Lisa et~al.}(2005)\citenamefont{Lisa, Pratt, Soltz, and
  Wiedemann}}]{Lisa:2005dd}
\bibinfo{author}{\bibfnamefont{M.~A.} \bibnamefont{Lisa}},
  \bibinfo{author}{\bibfnamefont{S.}~\bibnamefont{Pratt}},
  \bibinfo{author}{\bibfnamefont{R.}~\bibnamefont{Soltz}}, \bibnamefont{and}
  \bibinfo{author}{\bibfnamefont{U.}~\bibnamefont{Wiedemann}},
  \bibinfo{journal}{Ann. Rev. Nucl. Part. Sci.} \textbf{\bibinfo{volume}{55}},
  \bibinfo{pages}{357} (\bibinfo{year}{2005}), \eprint{nucl-ex/0505014}.

\bibitem[{\citenamefont{Collaboration et~al.}(2020)}]{ALICE:2020mfd}
\bibinfo{author}{\bibfnamefont{A.}~\bibnamefont{Collaboration}}
  \bibnamefont{et~al.} (\bibinfo{collaboration}{ALICE}),
  \bibinfo{journal}{Nature} \textbf{\bibinfo{volume}{588}},
  \bibinfo{pages}{232} (\bibinfo{year}{2020}), \bibinfo{note}{[Erratum: Nature
  590, E13 (2021)]}, \eprint{2005.11495}.

\bibitem[{ALI(2022)}]{ALICE:2022wwr}
 (\bibinfo{year}{2022}), \eprint{2211.02491}.

\bibitem[{\citenamefont{Fabbietti et~al.}(2021)\citenamefont{Fabbietti,
  Mantovani~Sarti, and Vazquez~Doce}}]{Fabbietti:2020bfg}
\bibinfo{author}{\bibfnamefont{L.}~\bibnamefont{Fabbietti}},
  \bibinfo{author}{\bibfnamefont{V.}~\bibnamefont{Mantovani~Sarti}},
  \bibnamefont{and}
  \bibinfo{author}{\bibfnamefont{O.}~\bibnamefont{Vazquez~Doce}},
  \bibinfo{journal}{Ann. Rev. Nucl. Part. Sci.} \textbf{\bibinfo{volume}{71}},
  \bibinfo{pages}{377} (\bibinfo{year}{2021}), \eprint{2012.09806}.

\bibitem[{\citenamefont{Acharya et~al.}(2019{\natexlab{a}})}]{ALICE:2018ysd}
\bibinfo{author}{\bibfnamefont{S.}~\bibnamefont{Acharya}} \bibnamefont{et~al.}
  (\bibinfo{collaboration}{ALICE}), \bibinfo{journal}{Phys. Rev. C}
  \textbf{\bibinfo{volume}{99}}, \bibinfo{pages}{024001}
  (\bibinfo{year}{2019}{\natexlab{a}}), \eprint{1805.12455}.

\bibitem[{\citenamefont{Acharya et~al.}(2020{\natexlab{a}})}]{ALICE:2019gcn}
\bibinfo{author}{\bibfnamefont{S.}~\bibnamefont{Acharya}} \bibnamefont{et~al.}
  (\bibinfo{collaboration}{ALICE}), \bibinfo{journal}{Phys. Rev. Lett.}
  \textbf{\bibinfo{volume}{124}}, \bibinfo{pages}{092301}
  (\bibinfo{year}{2020}{\natexlab{a}}), \eprint{1905.13470}.

\bibitem[{\citenamefont{Acharya et~al.}(2022{\natexlab{a}})}]{ALICE:2021njx}
\bibinfo{author}{\bibfnamefont{S.}~\bibnamefont{Acharya}} \bibnamefont{et~al.}
  (\bibinfo{collaboration}{ALICE}), \bibinfo{journal}{Phys. Lett. B}
  \textbf{\bibinfo{volume}{833}}, \bibinfo{pages}{137272}
  (\bibinfo{year}{2022}{\natexlab{a}}), \eprint{2104.04427}.

\bibitem[{\citenamefont{Acharya et~al.}(2020{\natexlab{b}})}]{ALICE:2019buq}
\bibinfo{author}{\bibfnamefont{S.}~\bibnamefont{Acharya}} \bibnamefont{et~al.}
  (\bibinfo{collaboration}{ALICE}), \bibinfo{journal}{Phys. Lett. B}
  \textbf{\bibinfo{volume}{805}}, \bibinfo{pages}{135419}
  (\bibinfo{year}{2020}{\natexlab{b}}), \eprint{1910.14407}.

\bibitem[{\citenamefont{Acharya et~al.}(2019{\natexlab{b}})}]{ALICE:2019eol}
\bibinfo{author}{\bibfnamefont{S.}~\bibnamefont{Acharya}} \bibnamefont{et~al.}
  (\bibinfo{collaboration}{ALICE}), \bibinfo{journal}{Phys. Lett. B}
  \textbf{\bibinfo{volume}{797}}, \bibinfo{pages}{134822}
  (\bibinfo{year}{2019}{\natexlab{b}}), \eprint{1905.07209}.

\bibitem[{\citenamefont{Acharya et~al.}(2019{\natexlab{c}})}]{ALICE:2019hdt}
\bibinfo{author}{\bibfnamefont{S.}~\bibnamefont{Acharya}} \bibnamefont{et~al.}
  (\bibinfo{collaboration}{ALICE}), \bibinfo{journal}{Phys. Rev. Lett.}
  \textbf{\bibinfo{volume}{123}}, \bibinfo{pages}{112002}
  (\bibinfo{year}{2019}{\natexlab{c}}), \eprint{1904.12198}.

\bibitem[{\citenamefont{Acharya et~al.}(2021)}]{ALICE:2021cpv}
\bibinfo{author}{\bibfnamefont{S.}~\bibnamefont{Acharya}} \bibnamefont{et~al.}
  (\bibinfo{collaboration}{ALICE}), \bibinfo{journal}{Phys. Rev. Lett.}
  \textbf{\bibinfo{volume}{127}}, \bibinfo{pages}{172301}
  (\bibinfo{year}{2021}), \eprint{2105.05578}.

\bibitem[{\citenamefont{Acharya et~al.}(2022{\natexlab{b}})}]{ALICE:2021cyj}
\bibinfo{author}{\bibfnamefont{S.}~\bibnamefont{Acharya}} \bibnamefont{et~al.}
  (\bibinfo{collaboration}{ALICE}), \bibinfo{journal}{Phys. Lett. B}
  \textbf{\bibinfo{volume}{829}}, \bibinfo{pages}{137060}
  (\bibinfo{year}{2022}{\natexlab{b}}), \eprint{2105.05190}.

\bibitem[{\citenamefont{Acharya et~al.}(2022{\natexlab{c}})}]{ALICE:2022enj}
\bibinfo{author}{\bibfnamefont{S.}~\bibnamefont{Acharya}} \bibnamefont{et~al.}
  (\bibinfo{collaboration}{ALICE}), \bibinfo{journal}{Phys. Rev. D}
  \textbf{\bibinfo{volume}{106}}, \bibinfo{pages}{052010}
  (\bibinfo{year}{2022}{\natexlab{c}}), \eprint{2201.05352}.

\bibitem[{\citenamefont{Koonin}(1977)}]{Koonin:1977fh}
\bibinfo{author}{\bibfnamefont{S.~E.} \bibnamefont{Koonin}},
  \bibinfo{journal}{Phys. Lett. B} \textbf{\bibinfo{volume}{70}},
  \bibinfo{pages}{43} (\bibinfo{year}{1977}).

\bibitem[{\citenamefont{Lednicky and Lyuboshits}(1981)}]{Lednicky:1981su}
\bibinfo{author}{\bibfnamefont{R.}~\bibnamefont{Lednicky}} \bibnamefont{and}
  \bibinfo{author}{\bibfnamefont{V.~L.} \bibnamefont{Lyuboshits}},
  \bibinfo{journal}{Yad. Fiz.} \textbf{\bibinfo{volume}{35}},
  \bibinfo{pages}{1316} (\bibinfo{year}{1981}).

\bibitem[{\citenamefont{Pratt}(1986)}]{Pratt:1986cc}
\bibinfo{author}{\bibfnamefont{S.}~\bibnamefont{Pratt}},
  \bibinfo{journal}{Phys. Rev. D} \textbf{\bibinfo{volume}{33}},
  \bibinfo{pages}{1314} (\bibinfo{year}{1986}).

\bibitem[{\citenamefont{Pratt et~al.}(1990)\citenamefont{Pratt, Csorgo, and
  Zimanyi}}]{Pratt:1990zq}
\bibinfo{author}{\bibfnamefont{S.}~\bibnamefont{Pratt}},
  \bibinfo{author}{\bibfnamefont{T.}~\bibnamefont{Csorgo}}, \bibnamefont{and}
  \bibinfo{author}{\bibfnamefont{J.}~\bibnamefont{Zimanyi}},
  \bibinfo{journal}{Phys. Rev. C} \textbf{\bibinfo{volume}{42}},
  \bibinfo{pages}{2646} (\bibinfo{year}{1990}).

\bibitem[{\citenamefont{Bauer et~al.}(1992)\citenamefont{Bauer, Gelbke, and
  Pratt}}]{Bauer:1992ffu}
\bibinfo{author}{\bibfnamefont{W.}~\bibnamefont{Bauer}},
  \bibinfo{author}{\bibfnamefont{C.~K.} \bibnamefont{Gelbke}},
  \bibnamefont{and} \bibinfo{author}{\bibfnamefont{S.}~\bibnamefont{Pratt}},
  \bibinfo{journal}{Ann. Rev. Nucl. Part. Sci.} \textbf{\bibinfo{volume}{42}},
  \bibinfo{pages}{77} (\bibinfo{year}{1992}).

\bibitem[{\citenamefont{Morita et~al.}(2015)\citenamefont{Morita, Furumoto, and
  Ohnishi}}]{Morita:2014kza}
\bibinfo{author}{\bibfnamefont{K.}~\bibnamefont{Morita}},
  \bibinfo{author}{\bibfnamefont{T.}~\bibnamefont{Furumoto}}, \bibnamefont{and}
  \bibinfo{author}{\bibfnamefont{A.}~\bibnamefont{Ohnishi}},
  \bibinfo{journal}{Phys. Rev. C} \textbf{\bibinfo{volume}{91}},
  \bibinfo{pages}{024916} (\bibinfo{year}{2015}), \eprint{1408.6682}.

\bibitem[{\citenamefont{Ohnishi et~al.}(2016)\citenamefont{Ohnishi, Morita,
  Miyahara, and Hyodo}}]{Ohnishi:2016elb}
\bibinfo{author}{\bibfnamefont{A.}~\bibnamefont{Ohnishi}},
  \bibinfo{author}{\bibfnamefont{K.}~\bibnamefont{Morita}},
  \bibinfo{author}{\bibfnamefont{K.}~\bibnamefont{Miyahara}}, \bibnamefont{and}
  \bibinfo{author}{\bibfnamefont{T.}~\bibnamefont{Hyodo}},
  \bibinfo{journal}{Nucl. Phys. A} \textbf{\bibinfo{volume}{954}},
  \bibinfo{pages}{294} (\bibinfo{year}{2016}), \eprint{1603.05761}.

\bibitem[{\citenamefont{Morita et~al.}(2016)\citenamefont{Morita, Ohnishi,
  Etminan, and Hatsuda}}]{Morita:2016auo}
\bibinfo{author}{\bibfnamefont{K.}~\bibnamefont{Morita}},
  \bibinfo{author}{\bibfnamefont{A.}~\bibnamefont{Ohnishi}},
  \bibinfo{author}{\bibfnamefont{F.}~\bibnamefont{Etminan}}, \bibnamefont{and}
  \bibinfo{author}{\bibfnamefont{T.}~\bibnamefont{Hatsuda}},
  \bibinfo{journal}{Phys. Rev. C} \textbf{\bibinfo{volume}{94}},
  \bibinfo{pages}{031901} (\bibinfo{year}{2016}), \bibinfo{note}{[Erratum:
  Phys.Rev.C 100, 069902 (2019)]}, \eprint{1605.06765}.

\bibitem[{\citenamefont{Hatsuda et~al.}(2017)\citenamefont{Hatsuda, Morita,
  Ohnishi, and Sasaki}}]{Hatsuda:2017uxk}
\bibinfo{author}{\bibfnamefont{T.}~\bibnamefont{Hatsuda}},
  \bibinfo{author}{\bibfnamefont{K.}~\bibnamefont{Morita}},
  \bibinfo{author}{\bibfnamefont{A.}~\bibnamefont{Ohnishi}}, \bibnamefont{and}
  \bibinfo{author}{\bibfnamefont{K.}~\bibnamefont{Sasaki}},
  \bibinfo{journal}{Nucl. Phys. A} \textbf{\bibinfo{volume}{967}},
  \bibinfo{pages}{856} (\bibinfo{year}{2017}), \eprint{1704.05225}.

\bibitem[{\citenamefont{Mihaylov et~al.}(2018)\citenamefont{Mihaylov,
  Mantovani~Sarti, Arnold, Fabbietti, Hohlweger, and
  Mathis}}]{Mihaylov:2018rva}
\bibinfo{author}{\bibfnamefont{D.~L.} \bibnamefont{Mihaylov}},
  \bibinfo{author}{\bibfnamefont{V.}~\bibnamefont{Mantovani~Sarti}},
  \bibinfo{author}{\bibfnamefont{O.~W.} \bibnamefont{Arnold}},
  \bibinfo{author}{\bibfnamefont{L.}~\bibnamefont{Fabbietti}},
  \bibinfo{author}{\bibfnamefont{B.}~\bibnamefont{Hohlweger}},
  \bibnamefont{and} \bibinfo{author}{\bibfnamefont{A.~M.}
  \bibnamefont{Mathis}}, \bibinfo{journal}{Eur. Phys. J. C}
  \textbf{\bibinfo{volume}{78}}, \bibinfo{pages}{394} (\bibinfo{year}{2018}),
  \eprint{1802.08481}.

\bibitem[{\citenamefont{Haidenbauer}(2019)}]{Haidenbauer:2018jvl}
\bibinfo{author}{\bibfnamefont{J.}~\bibnamefont{Haidenbauer}},
  \bibinfo{journal}{Nucl. Phys. A} \textbf{\bibinfo{volume}{981}},
  \bibinfo{pages}{1} (\bibinfo{year}{2019}), \eprint{1808.05049}.

\bibitem[{\citenamefont{Morita et~al.}(2020)\citenamefont{Morita, Gongyo,
  Hatsuda, Hyodo, Kamiya, and Ohnishi}}]{Morita:2019rph}
\bibinfo{author}{\bibfnamefont{K.}~\bibnamefont{Morita}},
  \bibinfo{author}{\bibfnamefont{S.}~\bibnamefont{Gongyo}},
  \bibinfo{author}{\bibfnamefont{T.}~\bibnamefont{Hatsuda}},
  \bibinfo{author}{\bibfnamefont{T.}~\bibnamefont{Hyodo}},
  \bibinfo{author}{\bibfnamefont{Y.}~\bibnamefont{Kamiya}}, \bibnamefont{and}
  \bibinfo{author}{\bibfnamefont{A.}~\bibnamefont{Ohnishi}},
  \bibinfo{journal}{Phys. Rev. C} \textbf{\bibinfo{volume}{101}},
  \bibinfo{pages}{015201} (\bibinfo{year}{2020}), \eprint{1908.05414}.

\bibitem[{\citenamefont{Kamiya et~al.}(2020)\citenamefont{Kamiya, Hyodo,
  Morita, Ohnishi, and Weise}}]{Kamiya:2019uiw}
\bibinfo{author}{\bibfnamefont{Y.}~\bibnamefont{Kamiya}},
  \bibinfo{author}{\bibfnamefont{T.}~\bibnamefont{Hyodo}},
  \bibinfo{author}{\bibfnamefont{K.}~\bibnamefont{Morita}},
  \bibinfo{author}{\bibfnamefont{A.}~\bibnamefont{Ohnishi}}, \bibnamefont{and}
  \bibinfo{author}{\bibfnamefont{W.}~\bibnamefont{Weise}},
  \bibinfo{journal}{Phys. Rev. Lett.} \textbf{\bibinfo{volume}{124}},
  \bibinfo{pages}{132501} (\bibinfo{year}{2020}), \eprint{1911.01041}.

\bibitem[{\citenamefont{Kamiya et~al.}(2022{\natexlab{a}})\citenamefont{Kamiya,
  Sasaki, Fukui, Hyodo, Morita, Ogata, Ohnishi, and Hatsuda}}]{Kamiya:2021hdb}
\bibinfo{author}{\bibfnamefont{Y.}~\bibnamefont{Kamiya}},
  \bibinfo{author}{\bibfnamefont{K.}~\bibnamefont{Sasaki}},
  \bibinfo{author}{\bibfnamefont{T.}~\bibnamefont{Fukui}},
  \bibinfo{author}{\bibfnamefont{T.}~\bibnamefont{Hyodo}},
  \bibinfo{author}{\bibfnamefont{K.}~\bibnamefont{Morita}},
  \bibinfo{author}{\bibfnamefont{K.}~\bibnamefont{Ogata}},
  \bibinfo{author}{\bibfnamefont{A.}~\bibnamefont{Ohnishi}}, \bibnamefont{and}
  \bibinfo{author}{\bibfnamefont{T.}~\bibnamefont{Hatsuda}},
  \bibinfo{journal}{Phys. Rev. C} \textbf{\bibinfo{volume}{105}},
  \bibinfo{pages}{014915} (\bibinfo{year}{2022}{\natexlab{a}}),
  \eprint{2108.09644}.

\bibitem[{\citenamefont{Kamiya et~al.}(2022{\natexlab{b}})\citenamefont{Kamiya,
  Hyodo, and Ohnishi}}]{Kamiya:2022thy}
\bibinfo{author}{\bibfnamefont{Y.}~\bibnamefont{Kamiya}},
  \bibinfo{author}{\bibfnamefont{T.}~\bibnamefont{Hyodo}}, \bibnamefont{and}
  \bibinfo{author}{\bibfnamefont{A.}~\bibnamefont{Ohnishi}},
  \bibinfo{journal}{Eur. Phys. J. A} \textbf{\bibinfo{volume}{58}},
  \bibinfo{pages}{131} (\bibinfo{year}{2022}{\natexlab{b}}),
  \eprint{2203.13814}.

\bibitem[{\citenamefont{Vidana et~al.}(2023)\citenamefont{Vidana, Feijoo,
  Albaladejo, Nieves, and Oset}}]{Vidana:2023olz}
\bibinfo{author}{\bibfnamefont{I.}~\bibnamefont{Vidana}},
  \bibinfo{author}{\bibfnamefont{A.}~\bibnamefont{Feijoo}},
  \bibinfo{author}{\bibfnamefont{M.}~\bibnamefont{Albaladejo}},
  \bibinfo{author}{\bibfnamefont{J.}~\bibnamefont{Nieves}}, \bibnamefont{and}
  \bibinfo{author}{\bibfnamefont{E.}~\bibnamefont{Oset}}
  (\bibinfo{year}{2023}), \eprint{2303.06079}.

\bibitem[{\citenamefont{Liu et~al.}(2023)\citenamefont{Liu, Lu, and
  Geng}}]{Liu:2023uly}
\bibinfo{author}{\bibfnamefont{Z.-W.} \bibnamefont{Liu}},
  \bibinfo{author}{\bibfnamefont{J.-X.} \bibnamefont{Lu}}, \bibnamefont{and}
  \bibinfo{author}{\bibfnamefont{L.-S.} \bibnamefont{Geng}},
  \bibinfo{journal}{Phys. Rev. D} \textbf{\bibinfo{volume}{107}},
  \bibinfo{pages}{074019} (\bibinfo{year}{2023}), \eprint{2302.01046}.

\bibitem[{\citenamefont{Albaladejo
  et~al.}(2023{\natexlab{a}})\citenamefont{Albaladejo, Nieves, and
  Ruiz-Arriola}}]{Albaladejo:2023pzq}
\bibinfo{author}{\bibfnamefont{M.}~\bibnamefont{Albaladejo}},
  \bibinfo{author}{\bibfnamefont{J.}~\bibnamefont{Nieves}}, \bibnamefont{and}
  \bibinfo{author}{\bibfnamefont{E.}~\bibnamefont{Ruiz-Arriola}},
  \bibinfo{journal}{Phys. Rev. D} \textbf{\bibinfo{volume}{108}},
  \bibinfo{pages}{014020} (\bibinfo{year}{2023}{\natexlab{a}}),
  \eprint{2304.03107}.

\bibitem[{\citenamefont{Torres-Rincon et~al.}(2023)\citenamefont{Torres-Rincon,
  Ramos, and Tolos}}]{Torres-Rincon:2023qll}
\bibinfo{author}{\bibfnamefont{J.~M.} \bibnamefont{Torres-Rincon}},
  \bibinfo{author}{\bibfnamefont{A.}~\bibnamefont{Ramos}}, \bibnamefont{and}
  \bibinfo{author}{\bibfnamefont{L.}~\bibnamefont{Tolos}}
  (\bibinfo{year}{2023}), \eprint{2307.02102}.

\bibitem[{\citenamefont{Ikeno et~al.}(2023)\citenamefont{Ikeno, Toledo, and
  Oset}}]{Ikeno:2023ojl}
\bibinfo{author}{\bibfnamefont{N.}~\bibnamefont{Ikeno}},
  \bibinfo{author}{\bibfnamefont{G.}~\bibnamefont{Toledo}}, \bibnamefont{and}
  \bibinfo{author}{\bibfnamefont{E.}~\bibnamefont{Oset}}
  (\bibinfo{year}{2023}), \eprint{2305.16431}.

\bibitem[{\citenamefont{Albaladejo
  et~al.}(2023{\natexlab{b}})\citenamefont{Albaladejo, Feijoo, Vida\~na,
  Nieves, and Oset}}]{Albaladejo:2023wmv}
\bibinfo{author}{\bibfnamefont{M.}~\bibnamefont{Albaladejo}},
  \bibinfo{author}{\bibfnamefont{A.}~\bibnamefont{Feijoo}},
  \bibinfo{author}{\bibfnamefont{I.}~\bibnamefont{Vida\~na}},
  \bibinfo{author}{\bibfnamefont{J.}~\bibnamefont{Nieves}}, \bibnamefont{and}
  \bibinfo{author}{\bibfnamefont{E.}~\bibnamefont{Oset}}
  (\bibinfo{year}{2023}{\natexlab{b}}), \eprint{2307.09873}.

\bibitem[{\citenamefont{Aaij et~al.}(2022{\natexlab{a}})}]{LHCb:2021vvq}
\bibinfo{author}{\bibfnamefont{R.}~\bibnamefont{Aaij}} \bibnamefont{et~al.}
  (\bibinfo{collaboration}{LHCb}), \bibinfo{journal}{Nature Phys.}
  \textbf{\bibinfo{volume}{18}}, \bibinfo{pages}{751}
  (\bibinfo{year}{2022}{\natexlab{a}}), \eprint{2109.01038}.

\bibitem[{\citenamefont{Aaij et~al.}(2022{\natexlab{b}})}]{LHCb:2021auc}
\bibinfo{author}{\bibfnamefont{R.}~\bibnamefont{Aaij}} \bibnamefont{et~al.}
  (\bibinfo{collaboration}{LHCb}), \bibinfo{journal}{Nature Commun.}
  \textbf{\bibinfo{volume}{13}}, \bibinfo{pages}{3351}
  (\bibinfo{year}{2022}{\natexlab{b}}), \eprint{2109.01056}.

\bibitem[{\citenamefont{Dai et~al.}(2022)\citenamefont{Dai, Oset, Feijoo,
  Molina, Roca, Torres, and Khemchandani}}]{Dai:2022ulk}
\bibinfo{author}{\bibfnamefont{L.~R.} \bibnamefont{Dai}},
  \bibinfo{author}{\bibfnamefont{E.}~\bibnamefont{Oset}},
  \bibinfo{author}{\bibfnamefont{A.}~\bibnamefont{Feijoo}},
  \bibinfo{author}{\bibfnamefont{R.}~\bibnamefont{Molina}},
  \bibinfo{author}{\bibfnamefont{L.}~\bibnamefont{Roca}},
  \bibinfo{author}{\bibfnamefont{A.~M.} \bibnamefont{Torres}},
  \bibnamefont{and} \bibinfo{author}{\bibfnamefont{K.~P.}
  \bibnamefont{Khemchandani}}, \bibinfo{journal}{Phys. Rev. D}
  \textbf{\bibinfo{volume}{105}}, \bibinfo{pages}{074017}
  (\bibinfo{year}{2022}), \bibinfo{note}{[Erratum: Phys.Rev.D 106, 099904
  (2022)]}, \eprint{2201.04840}.

\bibitem[{\citenamefont{Tornqvist}(1994)}]{Tornqvist:1993ng}
\bibinfo{author}{\bibfnamefont{N.~A.} \bibnamefont{Tornqvist}},
  \bibinfo{journal}{Z. Phys. C} \textbf{\bibinfo{volume}{61}},
  \bibinfo{pages}{525} (\bibinfo{year}{1994}), \eprint{hep-ph/9310247}.

\bibitem[{\citenamefont{Ohkoda et~al.}(2012)\citenamefont{Ohkoda, Yamaguchi,
  Yasui, Sudoh, and Hosaka}}]{Ohkoda:2012hv}
\bibinfo{author}{\bibfnamefont{S.}~\bibnamefont{Ohkoda}},
  \bibinfo{author}{\bibfnamefont{Y.}~\bibnamefont{Yamaguchi}},
  \bibinfo{author}{\bibfnamefont{S.}~\bibnamefont{Yasui}},
  \bibinfo{author}{\bibfnamefont{K.}~\bibnamefont{Sudoh}}, \bibnamefont{and}
  \bibinfo{author}{\bibfnamefont{A.}~\bibnamefont{Hosaka}},
  \bibinfo{journal}{Phys. Rev. D} \textbf{\bibinfo{volume}{86}},
  \bibinfo{pages}{034019} (\bibinfo{year}{2012}), \eprint{1202.0760}.

\bibitem[{\citenamefont{Li et~al.}(2013)\citenamefont{Li, Sun, Liu, and
  Zhu}}]{Li:2012ss}
\bibinfo{author}{\bibfnamefont{N.}~\bibnamefont{Li}},
  \bibinfo{author}{\bibfnamefont{Z.-F.} \bibnamefont{Sun}},
  \bibinfo{author}{\bibfnamefont{X.}~\bibnamefont{Liu}}, \bibnamefont{and}
  \bibinfo{author}{\bibfnamefont{S.-L.} \bibnamefont{Zhu}},
  \bibinfo{journal}{Phys. Rev. D} \textbf{\bibinfo{volume}{88}},
  \bibinfo{pages}{114008} (\bibinfo{year}{2013}), \eprint{1211.5007}.

\bibitem[{\citenamefont{Liu et~al.}(2020)\citenamefont{Liu, Xie, and
  Geng}}]{Liu:2020nil}
\bibinfo{author}{\bibfnamefont{M.-Z.} \bibnamefont{Liu}},
  \bibinfo{author}{\bibfnamefont{J.-J.} \bibnamefont{Xie}}, \bibnamefont{and}
  \bibinfo{author}{\bibfnamefont{L.-S.} \bibnamefont{Geng}},
  \bibinfo{journal}{Phys. Rev. D} \textbf{\bibinfo{volume}{102}},
  \bibinfo{pages}{091502} (\bibinfo{year}{2020}), \eprint{2008.07389}.

\bibitem[{\citenamefont{Manohar and Wise}(1993)}]{Manohar:1992nd}
\bibinfo{author}{\bibfnamefont{A.~V.} \bibnamefont{Manohar}} \bibnamefont{and}
  \bibinfo{author}{\bibfnamefont{M.~B.} \bibnamefont{Wise}},
  \bibinfo{journal}{Nucl. Phys. B} \textbf{\bibinfo{volume}{399}},
  \bibinfo{pages}{17} (\bibinfo{year}{1993}), \eprint{hep-ph/9212236}.

\bibitem[{\citenamefont{Zhao et~al.}(2022)\citenamefont{Zhao, Wang, Wang, and
  Guo}}]{Zhao:2021cvg}
\bibinfo{author}{\bibfnamefont{M.-J.} \bibnamefont{Zhao}},
  \bibinfo{author}{\bibfnamefont{Z.-Y.} \bibnamefont{Wang}},
  \bibinfo{author}{\bibfnamefont{C.}~\bibnamefont{Wang}}, \bibnamefont{and}
  \bibinfo{author}{\bibfnamefont{X.-H.} \bibnamefont{Guo}},
  \bibinfo{journal}{Phys. Rev. D} \textbf{\bibinfo{volume}{105}},
  \bibinfo{pages}{096016} (\bibinfo{year}{2022}), \eprint{2112.12633}.

\bibitem[{\citenamefont{Ke et~al.}(2022)\citenamefont{Ke, Liu, and
  Li}}]{Ke:2021rxd}
\bibinfo{author}{\bibfnamefont{H.-W.} \bibnamefont{Ke}},
  \bibinfo{author}{\bibfnamefont{X.-H.} \bibnamefont{Liu}}, \bibnamefont{and}
  \bibinfo{author}{\bibfnamefont{X.-Q.} \bibnamefont{Li}},
  \bibinfo{journal}{Eur. Phys. J. C} \textbf{\bibinfo{volume}{82}},
  \bibinfo{pages}{144} (\bibinfo{year}{2022}), \eprint{2112.14142}.

\bibitem[{\citenamefont{Yang et~al.}(2009)\citenamefont{Yang, Deng, Ping, and
  Goldman}}]{Yang:2009zzp}
\bibinfo{author}{\bibfnamefont{Y.}~\bibnamefont{Yang}},
  \bibinfo{author}{\bibfnamefont{C.}~\bibnamefont{Deng}},
  \bibinfo{author}{\bibfnamefont{J.}~\bibnamefont{Ping}}, \bibnamefont{and}
  \bibinfo{author}{\bibfnamefont{T.}~\bibnamefont{Goldman}},
  \bibinfo{journal}{Phys. Rev. D} \textbf{\bibinfo{volume}{80}},
  \bibinfo{pages}{114023} (\bibinfo{year}{2009}).

\bibitem[{\citenamefont{Barnes et~al.}(1999)\citenamefont{Barnes, Black, Dean,
  and Swanson}}]{Barnes:1999hs}
\bibinfo{author}{\bibfnamefont{T.}~\bibnamefont{Barnes}},
  \bibinfo{author}{\bibfnamefont{N.}~\bibnamefont{Black}},
  \bibinfo{author}{\bibfnamefont{D.~J.} \bibnamefont{Dean}}, \bibnamefont{and}
  \bibinfo{author}{\bibfnamefont{E.~S.} \bibnamefont{Swanson}},
  \bibinfo{journal}{Phys. Rev. C} \textbf{\bibinfo{volume}{60}},
  \bibinfo{pages}{045202} (\bibinfo{year}{1999}), \eprint{nucl-th/9902068}.

\bibitem[{\citenamefont{Michael and Pennanen}(1999)}]{Michael:1999nq}
\bibinfo{author}{\bibfnamefont{C.}~\bibnamefont{Michael}} \bibnamefont{and}
  \bibinfo{author}{\bibfnamefont{P.}~\bibnamefont{Pennanen}}
  (\bibinfo{collaboration}{UKQCD}), \bibinfo{journal}{Phys. Rev. D}
  \textbf{\bibinfo{volume}{60}}, \bibinfo{pages}{054012}
  (\bibinfo{year}{1999}), \eprint{hep-lat/9901007}.

\bibitem[{\citenamefont{Detmold et~al.}(2007)\citenamefont{Detmold, Orginos,
  and Savage}}]{Detmold:2007wk}
\bibinfo{author}{\bibfnamefont{W.}~\bibnamefont{Detmold}},
  \bibinfo{author}{\bibfnamefont{K.}~\bibnamefont{Orginos}}, \bibnamefont{and}
  \bibinfo{author}{\bibfnamefont{M.~J.} \bibnamefont{Savage}},
  \bibinfo{journal}{Phys. Rev. D} \textbf{\bibinfo{volume}{76}},
  \bibinfo{pages}{114503} (\bibinfo{year}{2007}), \eprint{hep-lat/0703009}.

\bibitem[{\citenamefont{Brown and Orginos}(2012)}]{Brown:2012tm}
\bibinfo{author}{\bibfnamefont{Z.~S.} \bibnamefont{Brown}} \bibnamefont{and}
  \bibinfo{author}{\bibfnamefont{K.}~\bibnamefont{Orginos}},
  \bibinfo{journal}{Phys. Rev. D} \textbf{\bibinfo{volume}{86}},
  \bibinfo{pages}{114506} (\bibinfo{year}{2012}), \eprint{1210.1953}.

\bibitem[{\citenamefont{Bicudo and Wagner}(2013)}]{Bicudo:2012qt}
\bibinfo{author}{\bibfnamefont{P.}~\bibnamefont{Bicudo}} \bibnamefont{and}
  \bibinfo{author}{\bibfnamefont{M.}~\bibnamefont{Wagner}}
  (\bibinfo{collaboration}{European Twisted Mass}), \bibinfo{journal}{Phys.
  Rev. D} \textbf{\bibinfo{volume}{87}}, \bibinfo{pages}{114511}
  (\bibinfo{year}{2013}), \eprint{1209.6274}.

\bibitem[{\citenamefont{Bicudo et~al.}(2016)\citenamefont{Bicudo, Cichy,
  Peters, and Wagner}}]{Bicudo:2015kna}
\bibinfo{author}{\bibfnamefont{P.}~\bibnamefont{Bicudo}},
  \bibinfo{author}{\bibfnamefont{K.}~\bibnamefont{Cichy}},
  \bibinfo{author}{\bibfnamefont{A.}~\bibnamefont{Peters}}, \bibnamefont{and}
  \bibinfo{author}{\bibfnamefont{M.}~\bibnamefont{Wagner}},
  \bibinfo{journal}{Phys. Rev. D} \textbf{\bibinfo{volume}{93}},
  \bibinfo{pages}{034501} (\bibinfo{year}{2016}), \eprint{1510.03441}.

\bibitem[{\citenamefont{Carlson et~al.}(1988)\citenamefont{Carlson, Heller, and
  Tjon}}]{Carlson:1987hh}
\bibinfo{author}{\bibfnamefont{J.}~\bibnamefont{Carlson}},
  \bibinfo{author}{\bibfnamefont{L.}~\bibnamefont{Heller}}, \bibnamefont{and}
  \bibinfo{author}{\bibfnamefont{J.~A.} \bibnamefont{Tjon}},
  \bibinfo{journal}{Phys. Rev. D} \textbf{\bibinfo{volume}{37}},
  \bibinfo{pages}{744} (\bibinfo{year}{1988}).

\bibitem[{\citenamefont{Bicudo et~al.}(2017)\citenamefont{Bicudo, Scheunert,
  and Wagner}}]{Bicudo:2016ooe}
\bibinfo{author}{\bibfnamefont{P.}~\bibnamefont{Bicudo}},
  \bibinfo{author}{\bibfnamefont{J.}~\bibnamefont{Scheunert}},
  \bibnamefont{and} \bibinfo{author}{\bibfnamefont{M.}~\bibnamefont{Wagner}},
  \bibinfo{journal}{Phys. Rev. D} \textbf{\bibinfo{volume}{95}},
  \bibinfo{pages}{034502} (\bibinfo{year}{2017}), \eprint{1612.02758}.

\bibitem[{\citenamefont{Sanchez~Sanchez
  et~al.}(2018)\citenamefont{Sanchez~Sanchez, Geng, Lu, Hyodo, and
  Valderrama}}]{SanchezSanchez:2017xtl}
\bibinfo{author}{\bibfnamefont{M.}~\bibnamefont{Sanchez~Sanchez}},
  \bibinfo{author}{\bibfnamefont{L.-S.} \bibnamefont{Geng}},
  \bibinfo{author}{\bibfnamefont{J.-X.} \bibnamefont{Lu}},
  \bibinfo{author}{\bibfnamefont{T.}~\bibnamefont{Hyodo}}, \bibnamefont{and}
  \bibinfo{author}{\bibfnamefont{M.~P.} \bibnamefont{Valderrama}},
  \bibinfo{journal}{Phys. Rev. D} \textbf{\bibinfo{volume}{98}},
  \bibinfo{pages}{054001} (\bibinfo{year}{2018}), \eprint{1707.03802}.

\bibitem[{\citenamefont{Wang et~al.}(2019)\citenamefont{Wang, Liu, and
  Liu}}]{Wang:2018atz}
\bibinfo{author}{\bibfnamefont{B.}~\bibnamefont{Wang}},
  \bibinfo{author}{\bibfnamefont{Z.-W.} \bibnamefont{Liu}}, \bibnamefont{and}
  \bibinfo{author}{\bibfnamefont{X.}~\bibnamefont{Liu}},
  \bibinfo{journal}{Phys. Rev. D} \textbf{\bibinfo{volume}{99}},
  \bibinfo{pages}{036007} (\bibinfo{year}{2019}), \eprint{1812.04457}.

\bibitem[{\citenamefont{Yu et~al.}(2020)\citenamefont{Yu, Zhou, Chen, and
  Xiao}}]{Yu:2019sxx}
\bibinfo{author}{\bibfnamefont{M.-T.} \bibnamefont{Yu}},
  \bibinfo{author}{\bibfnamefont{Z.-Y.} \bibnamefont{Zhou}},
  \bibinfo{author}{\bibfnamefont{D.-Y.} \bibnamefont{Chen}}, \bibnamefont{and}
  \bibinfo{author}{\bibfnamefont{Z.}~\bibnamefont{Xiao}},
  \bibinfo{journal}{Phys. Rev. D} \textbf{\bibinfo{volume}{101}},
  \bibinfo{pages}{074027} (\bibinfo{year}{2020}), \eprint{1912.07348}.

\bibitem[{\citenamefont{Ding et~al.}(2020)\citenamefont{Ding, Jiang, and
  He}}]{Ding:2020dio}
\bibinfo{author}{\bibfnamefont{Z.-M.} \bibnamefont{Ding}},
  \bibinfo{author}{\bibfnamefont{H.-Y.} \bibnamefont{Jiang}}, \bibnamefont{and}
  \bibinfo{author}{\bibfnamefont{J.}~\bibnamefont{He}}, \bibinfo{journal}{Eur.
  Phys. J. C} \textbf{\bibinfo{volume}{80}}, \bibinfo{pages}{1179}
  (\bibinfo{year}{2020}), \eprint{2011.04980}.

\bibitem[{\citenamefont{Ding et~al.}(2021)\citenamefont{Ding, Jiang, Song, and
  He}}]{Ding:2021igr}
\bibinfo{author}{\bibfnamefont{Z.-M.} \bibnamefont{Ding}},
  \bibinfo{author}{\bibfnamefont{H.-Y.} \bibnamefont{Jiang}},
  \bibinfo{author}{\bibfnamefont{D.}~\bibnamefont{Song}}, \bibnamefont{and}
  \bibinfo{author}{\bibfnamefont{J.}~\bibnamefont{He}}, \bibinfo{journal}{Eur.
  Phys. J. C} \textbf{\bibinfo{volume}{81}}, \bibinfo{pages}{732}
  (\bibinfo{year}{2021}), \eprint{2107.00855}.

\bibitem[{\citenamefont{Meng et~al.}(2021)\citenamefont{Meng, Hiyama, Hosaka,
  Oka, Gubler, Can, Takahashi, and Zong}}]{Meng:2020knc}
\bibinfo{author}{\bibfnamefont{Q.}~\bibnamefont{Meng}},
  \bibinfo{author}{\bibfnamefont{E.}~\bibnamefont{Hiyama}},
  \bibinfo{author}{\bibfnamefont{A.}~\bibnamefont{Hosaka}},
  \bibinfo{author}{\bibfnamefont{M.}~\bibnamefont{Oka}},
  \bibinfo{author}{\bibfnamefont{P.}~\bibnamefont{Gubler}},
  \bibinfo{author}{\bibfnamefont{K.~U.} \bibnamefont{Can}},
  \bibinfo{author}{\bibfnamefont{T.~T.} \bibnamefont{Takahashi}},
  \bibnamefont{and} \bibinfo{author}{\bibfnamefont{H.~S.} \bibnamefont{Zong}},
  \bibinfo{journal}{Phys. Lett. B} \textbf{\bibinfo{volume}{814}},
  \bibinfo{pages}{136095} (\bibinfo{year}{2021}), \eprint{2009.14493}.

\bibitem[{\citenamefont{Abreu}(2022)}]{Abreu:2022sra}
\bibinfo{author}{\bibfnamefont{L.~M.} \bibnamefont{Abreu}},
  \bibinfo{journal}{Nucl. Phys. B} \textbf{\bibinfo{volume}{985}},
  \bibinfo{pages}{115994} (\bibinfo{year}{2022}), \eprint{2206.01166}.

\bibitem[{\citenamefont{Ader et~al.}(1982)\citenamefont{Ader, Richard, and
  Taxil}}]{Ader:1981db}
\bibinfo{author}{\bibfnamefont{J.~P.} \bibnamefont{Ader}},
  \bibinfo{author}{\bibfnamefont{J.~M.} \bibnamefont{Richard}},
  \bibnamefont{and} \bibinfo{author}{\bibfnamefont{P.}~\bibnamefont{Taxil}},
  \bibinfo{journal}{Phys. Rev. D} \textbf{\bibinfo{volume}{25}},
  \bibinfo{pages}{2370} (\bibinfo{year}{1982}).

\bibitem[{\citenamefont{Hern\'andez et~al.}(2020)\citenamefont{Hern\'andez,
  Vijande, Valcarce, and Richard}}]{Hernandez:2019eox}
\bibinfo{author}{\bibfnamefont{E.}~\bibnamefont{Hern\'andez}},
  \bibinfo{author}{\bibfnamefont{J.}~\bibnamefont{Vijande}},
  \bibinfo{author}{\bibfnamefont{A.}~\bibnamefont{Valcarce}}, \bibnamefont{and}
  \bibinfo{author}{\bibfnamefont{J.-M.} \bibnamefont{Richard}},
  \bibinfo{journal}{Phys. Lett. B} \textbf{\bibinfo{volume}{800}},
  \bibinfo{pages}{135073} (\bibinfo{year}{2020}), \eprint{1910.13394}.

\bibitem[{\citenamefont{Francis et~al.}(2017)\citenamefont{Francis, Hudspith,
  Lewis, and Maltman}}]{Francis:2016hui}
\bibinfo{author}{\bibfnamefont{A.}~\bibnamefont{Francis}},
  \bibinfo{author}{\bibfnamefont{R.~J.} \bibnamefont{Hudspith}},
  \bibinfo{author}{\bibfnamefont{R.}~\bibnamefont{Lewis}}, \bibnamefont{and}
  \bibinfo{author}{\bibfnamefont{K.}~\bibnamefont{Maltman}},
  \bibinfo{journal}{Phys. Rev. Lett.} \textbf{\bibinfo{volume}{118}},
  \bibinfo{pages}{142001} (\bibinfo{year}{2017}), \eprint{1607.05214}.

\bibitem[{\citenamefont{Junnarkar et~al.}(2019)\citenamefont{Junnarkar, Mathur,
  and Padmanath}}]{Junnarkar:2018twb}
\bibinfo{author}{\bibfnamefont{P.}~\bibnamefont{Junnarkar}},
  \bibinfo{author}{\bibfnamefont{N.}~\bibnamefont{Mathur}}, \bibnamefont{and}
  \bibinfo{author}{\bibfnamefont{M.}~\bibnamefont{Padmanath}},
  \bibinfo{journal}{Phys. Rev. D} \textbf{\bibinfo{volume}{99}},
  \bibinfo{pages}{034507} (\bibinfo{year}{2019}), \eprint{1810.12285}.

\bibitem[{\citenamefont{Leskovec et~al.}(2019)\citenamefont{Leskovec, Meinel,
  Pflaumer, and Wagner}}]{Leskovec:2019ioa}
\bibinfo{author}{\bibfnamefont{L.}~\bibnamefont{Leskovec}},
  \bibinfo{author}{\bibfnamefont{S.}~\bibnamefont{Meinel}},
  \bibinfo{author}{\bibfnamefont{M.}~\bibnamefont{Pflaumer}}, \bibnamefont{and}
  \bibinfo{author}{\bibfnamefont{M.}~\bibnamefont{Wagner}},
  \bibinfo{journal}{Phys. Rev. D} \textbf{\bibinfo{volume}{100}},
  \bibinfo{pages}{014503} (\bibinfo{year}{2019}), \eprint{1904.04197}.

\bibitem[{\citenamefont{Mohanta and Basak}(2020)}]{Mohanta:2020eed}
\bibinfo{author}{\bibfnamefont{P.}~\bibnamefont{Mohanta}} \bibnamefont{and}
  \bibinfo{author}{\bibfnamefont{S.}~\bibnamefont{Basak}},
  \bibinfo{journal}{Phys. Rev. D} \textbf{\bibinfo{volume}{102}},
  \bibinfo{pages}{094516} (\bibinfo{year}{2020}), \eprint{2008.11146}.

\bibitem[{\citenamefont{Hudspith and Mohler}(2023)}]{Hudspith:2023loy}
\bibinfo{author}{\bibfnamefont{R.~J.} \bibnamefont{Hudspith}} \bibnamefont{and}
  \bibinfo{author}{\bibfnamefont{D.}~\bibnamefont{Mohler}},
  \bibinfo{journal}{Phys. Rev. D} \textbf{\bibinfo{volume}{107}},
  \bibinfo{pages}{114510} (\bibinfo{year}{2023}), \eprint{2303.17295}.

\bibitem[{\citenamefont{Aoki et~al.}(2023)\citenamefont{Aoki, Aoki, and
  Inoue}}]{Aoki:2023nzp}
\bibinfo{author}{\bibfnamefont{T.}~\bibnamefont{Aoki}},
  \bibinfo{author}{\bibfnamefont{S.}~\bibnamefont{Aoki}}, \bibnamefont{and}
  \bibinfo{author}{\bibfnamefont{T.}~\bibnamefont{Inoue}},
  \bibinfo{journal}{Phys. Rev. D} \textbf{\bibinfo{volume}{108}},
  \bibinfo{pages}{054502} (\bibinfo{year}{2023}), \eprint{2306.03565}.

\bibitem[{\citenamefont{Bando et~al.}(1988)\citenamefont{Bando, Kugo, and
  Yamawaki}}]{Bando:1987br}
\bibinfo{author}{\bibfnamefont{M.}~\bibnamefont{Bando}},
  \bibinfo{author}{\bibfnamefont{T.}~\bibnamefont{Kugo}}, \bibnamefont{and}
  \bibinfo{author}{\bibfnamefont{K.}~\bibnamefont{Yamawaki}},
  \bibinfo{journal}{Phys. Rept.} \textbf{\bibinfo{volume}{164}},
  \bibinfo{pages}{217} (\bibinfo{year}{1988}).

\bibitem[{\citenamefont{Harada and Yamawaki}(2003)}]{Harada:2003jx}
\bibinfo{author}{\bibfnamefont{M.}~\bibnamefont{Harada}} \bibnamefont{and}
  \bibinfo{author}{\bibfnamefont{K.}~\bibnamefont{Yamawaki}},
  \bibinfo{journal}{Phys. Rept.} \textbf{\bibinfo{volume}{381}},
  \bibinfo{pages}{1} (\bibinfo{year}{2003}), \eprint{hep-ph/0302103}.

\bibitem[{\citenamefont{Meissner}(1988)}]{Meissner:1987ge}
\bibinfo{author}{\bibfnamefont{U.~G.} \bibnamefont{Meissner}},
  \bibinfo{journal}{Phys. Rept.} \textbf{\bibinfo{volume}{161}},
  \bibinfo{pages}{213} (\bibinfo{year}{1988}).

\bibitem[{\citenamefont{Nagahiro et~al.}(2009)\citenamefont{Nagahiro, Roca,
  Hosaka, and Oset}}]{Nagahiro:2008cv}
\bibinfo{author}{\bibfnamefont{H.}~\bibnamefont{Nagahiro}},
  \bibinfo{author}{\bibfnamefont{L.}~\bibnamefont{Roca}},
  \bibinfo{author}{\bibfnamefont{A.}~\bibnamefont{Hosaka}}, \bibnamefont{and}
  \bibinfo{author}{\bibfnamefont{E.}~\bibnamefont{Oset}},
  \bibinfo{journal}{Phys. Rev. D} \textbf{\bibinfo{volume}{79}},
  \bibinfo{pages}{014015} (\bibinfo{year}{2009}), \eprint{0809.0943}.

\bibitem[{\citenamefont{Oller and Meissner}(2001)}]{Oller:2000fj}
\bibinfo{author}{\bibfnamefont{J.~A.} \bibnamefont{Oller}} \bibnamefont{and}
  \bibinfo{author}{\bibfnamefont{U.~G.} \bibnamefont{Meissner}},
  \bibinfo{journal}{Phys. Lett. B} \textbf{\bibinfo{volume}{500}},
  \bibinfo{pages}{263} (\bibinfo{year}{2001}), \eprint{hep-ph/0011146}.

\bibitem[{\citenamefont{Dai et~al.}(2023)\citenamefont{Dai, Abreu, Feijoo, and
  Oset}}]{Dai:2023cyo}
\bibinfo{author}{\bibfnamefont{L.~R.} \bibnamefont{Dai}},
  \bibinfo{author}{\bibfnamefont{L.~M.} \bibnamefont{Abreu}},
  \bibinfo{author}{\bibfnamefont{A.}~\bibnamefont{Feijoo}}, \bibnamefont{and}
  \bibinfo{author}{\bibfnamefont{E.}~\bibnamefont{Oset}},
  \bibinfo{journal}{Eur. Phys. J. C} \textbf{\bibinfo{volume}{83}},
  \bibinfo{pages}{983} (\bibinfo{year}{2023}), \eprint{2304.01870}.

\bibitem[{\citenamefont{Hyodo}(2013)}]{Hyodo:2013nka}
\bibinfo{author}{\bibfnamefont{T.}~\bibnamefont{Hyodo}}, \bibinfo{journal}{Int.
  J. Mod. Phys. A} \textbf{\bibinfo{volume}{28}}, \bibinfo{pages}{1330045}
  (\bibinfo{year}{2013}), \eprint{1310.1176}.

\bibitem[{\citenamefont{Aceti et~al.}(2014)\citenamefont{Aceti, Dai, Geng,
  Oset, and Zhang}}]{Aceti:2014ala}
\bibinfo{author}{\bibfnamefont{F.}~\bibnamefont{Aceti}},
  \bibinfo{author}{\bibfnamefont{L.~R.} \bibnamefont{Dai}},
  \bibinfo{author}{\bibfnamefont{L.~S.} \bibnamefont{Geng}},
  \bibinfo{author}{\bibfnamefont{E.}~\bibnamefont{Oset}}, \bibnamefont{and}
  \bibinfo{author}{\bibfnamefont{Y.}~\bibnamefont{Zhang}},
  \bibinfo{journal}{Eur. Phys. J. A} \textbf{\bibinfo{volume}{50}},
  \bibinfo{pages}{57} (\bibinfo{year}{2014}), \eprint{1301.2554}.

\bibitem[{\citenamefont{Press et~al.}(1992)\citenamefont{Press, Teukolsky,
  Vetterling, and Flannery}}]{PresTeukVettFlan92}
\bibinfo{author}{\bibfnamefont{W.~H.} \bibnamefont{Press}},
  \bibinfo{author}{\bibfnamefont{S.~A.} \bibnamefont{Teukolsky}},
  \bibinfo{author}{\bibfnamefont{W.~T.} \bibnamefont{Vetterling}},
  \bibnamefont{and} \bibinfo{author}{\bibfnamefont{B.~P.}
  \bibnamefont{Flannery}}, \emph{\bibinfo{title}{Numerical Recipes in C}}
  (\bibinfo{publisher}{Cambridge University Press},
  \bibinfo{address}{Cambridge, USA}, \bibinfo{year}{1992}),
  \bibinfo{edition}{2nd} ed.

\bibitem[{\citenamefont{Efron and Tibshirani}(1986)}]{Efron:1986hys}
\bibinfo{author}{\bibfnamefont{B.}~\bibnamefont{Efron}} \bibnamefont{and}
  \bibinfo{author}{\bibfnamefont{R.}~\bibnamefont{Tibshirani}},
  \bibinfo{journal}{Statist. Sci.} \textbf{\bibinfo{volume}{57}},
  \bibinfo{pages}{54} (\bibinfo{year}{1986}).

\bibitem[{\citenamefont{Albaladejo et~al.}(2016)\citenamefont{Albaladejo, Jido,
  Nieves, and Oset}}]{Albaladejo:2016hae}
\bibinfo{author}{\bibfnamefont{M.}~\bibnamefont{Albaladejo}},
  \bibinfo{author}{\bibfnamefont{D.}~\bibnamefont{Jido}},
  \bibinfo{author}{\bibfnamefont{J.}~\bibnamefont{Nieves}}, \bibnamefont{and}
  \bibinfo{author}{\bibfnamefont{E.}~\bibnamefont{Oset}},
  \bibinfo{journal}{Eur. Phys. J. C} \textbf{\bibinfo{volume}{76}},
  \bibinfo{pages}{300} (\bibinfo{year}{2016}), \eprint{1604.01193}.

\bibitem[{\citenamefont{Oller et~al.}(1999)\citenamefont{Oller, Oset, and
  Pelaez}}]{Oller:1998hw}
\bibinfo{author}{\bibfnamefont{J.~A.} \bibnamefont{Oller}},
  \bibinfo{author}{\bibfnamefont{E.}~\bibnamefont{Oset}}, \bibnamefont{and}
  \bibinfo{author}{\bibfnamefont{J.~R.} \bibnamefont{Pelaez}},
  \bibinfo{journal}{Phys. Rev. D} \textbf{\bibinfo{volume}{59}},
  \bibinfo{pages}{074001} (\bibinfo{year}{1999}), \bibinfo{note}{[Erratum:
  Phys.Rev.D 60, 099906 (1999), Erratum: Phys.Rev.D 75, 099903 (2007)]},
  \eprint{hep-ph/9804209}.

\end{thebibliography}

\end{document}